\documentclass[letterpaper]{article} % DO NOT CHANGE THIS
\usepackage{aaai25}  % DO NOT CHANGE THIS
\usepackage{times}  % DO NOT CHANGE THIS
\usepackage{helvet}  % DO NOT CHANGE THIS
\usepackage{courier}  % DO NOT CHANGE THIS
\usepackage[hyphens]{url}  % DO NOT CHANGE THIS
\usepackage{graphicx} % DO NOT CHANGE THIS
\usepackage{natbib}  % DO NOT CHANGE THIS AND DO NOT ADD ANY OPTIONS TO IT
\usepackage{caption} % DO NOT CHANGE THIS AND DO NOT ADD ANY OPTIONS TO IT
\usepackage{algorithm}
\usepackage{algorithmic}

\usepackage{newfloat}
\usepackage{listings}
\DeclareCaptionStyle{ruled}{labelfont=normalfont,labelsep=colon,strut=off} % DO NOT CHANGE THIS
\floatstyle{ruled}
\newfloat{listing}{tb}{lst}{}
\floatname{listing}{Listing}
\usepackage{amsmath}
\usepackage{amssymb}
\usepackage{multirow}
\usepackage{cleveref}

\title{Sybil Detection using Graph Neural Networks\footnote{Under review}}
\author{
    Stuart Heeb\equalcontrib,
    Andreas Plesner\equalcontrib\footnote{Corresponding author \texttt{aplesner@ethz.ch}.},
    Roger Wattenhofer
}
\affiliations{
    ETH Zurich\\
}

\begin{document}

\maketitle

\begin{abstract}
This paper presents \textsc{SybilGAT}, a novel approach to Sybil detection in social networks using Graph Attention Networks (GATs). Traditional methods for Sybil detection primarily leverage structural properties of networks; however, they tend to struggle with a large number of attack edges and are often unable to simultaneously utilize both known Sybil and honest nodes. Our proposed method addresses these limitations by dynamically assigning attention weights to different nodes during aggregations, enhancing detection performance. We conducted extensive experiments in various scenarios, including pretraining in sampled subgraphs, synthetic networks, and networks under targeted attacks. The results show that \textsc{SybilGAT} significantly outperforms the state-of-the-art algorithms, particularly in scenarios with high attack complexity and when the number of attack edges increases. Our approach shows robust performance across different network models and sizes, even as the detection task becomes more challenging. 
We successfully applied the model to a real-world Twitter graph with more than 269k nodes and 6.8M edges.
The flexibility and generalizability of \textsc{SybilGAT} make it a promising tool to defend against Sybil attacks in online social networks with only structural information.
\end{abstract}

% Uncomment the following to link to your code, datasets, an extended version or similar.
%
% \begin{links}
%     \link{Code}{https://aaai.org/example/code}
%     \link{Datasets}{https://aaai.org/example/datasets}
%     \link{Extended version}{https://aaai.org/example/extended-version}
% \end{links}

\section{Introduction}

Online social networks have become a central part of modern digital life, connecting billions of users worldwide. However, their open nature and vast scale make them vulnerable to various security threats, with Sybil attacks being a prevalent and challenging attack to detect. In Sybil attacks, malicious entities create fake accounts to manipulate the network, spread misinformation, or gain undue influence.
Thus, detecting Sybil accounts is crucial for maintaining the integrity and trustworthiness of social networks. Effective Sybil detection can prevent the spread of fake news, protect users from scams, and ensure a fair distribution of resources and influence within the network. However, as attackers become more sophisticated, traditional detection methods increasingly fail to perform their task.

The rise of modern generative AI has allowed the creation of user features, images, and texts that closely mimic genuine human activity, leaving the structural features of networks as the most reliable indicators for Sybil detection. Although current approaches, such as random walks and loopy belief propagation, have demonstrated effectiveness, they also exhibit significant drawbacks. These techniques often suffer from issues like label noise, inability to concurrently leverage information from both known Sybil and honest nodes, and vulnerability to sophisticated attack strategies.

In recent years, Graph Neural Networks (GNNs) have emerged as a powerful tool for learning on graph-structured data. GNNs can capture complex patterns and relationships within networks, making them a promising approach to Sybil detection. However, their application to this specific security challenge remains underexplored.
In this work, we introduce \textsc{SybilGAT}, a novel approach to Sybil detection that leverages Graph Attention Networks (GATs), a specific type of GNN. \textsc{SybilGAT} addresses the limitations of existing methods by dynamically assigning attention weights to different nodes during the aggregation process, allowing it to focus on the most relevant information for detection.
We have conducted extensive experiments to evaluate the performance of \textsc{SybilGAT} in various scenarios. These include pre-training on sampled subgraphs, testing on synthetic networks of different sizes and structures, and assessing robustness against targeted attacks. Our experiments use real-world datasets, such as Twitter and Facebook networks, and synthetically generated graphs based on well-known models such as Barabási-Albert and Power law graphs.

The results show that \textsc{SybilGAT} significantly outperforms the state-of-the-art algorithms, particularly in scenarios with high attack complexity and many attack edges. Our approach shows robust performance across different network models and sizes, even as the detection task becomes more challenging. 

% Through this work, we have shown that Graph Attention Networks can be effectively applied to the problem of Sybil detection, offering a more flexible and generalizable solution than traditional methods. The success of \textsc{SybilGAT} opens up new avenues for research in applying advanced graph learning techniques to network security challenges.

\section{Related Work}\label{sec:related-work}

\paragraph{Structure-based methods}

The majority of previous work considered for this research were structure-based methods, meaning that the only features available for Sybil detection are the graph structure and a set of known (honest and Sybil) nodes. Many methods heavily use the \emph{homophily assumption}, which states that nodes connected by an edge tend to share the same label. Based on this assumption, the labels of a few known nodes are propagated through the social network. This is generally done using random walks (RW)
~\cite{sybilguard,sybillimit,sybilinfer,sybilrank,integro,sybilwalk} 
or loopy belief propagation (LBP)
~\cite{sybilbelief,sybilfuse,gang,sybilscar-original,sybilscar}
.

% RW based

\textsc{SybilGuard}~\cite{sybilguard} and \textsc{SybilLimit}~\cite{sybillimit} assume that it is more likely for a random walk starting from a known honest node to reach other honest nodes than Sybil nodes, and vice versa. They use the same length of random walks for all nodes. While \textsc{SybilGuard} accepts $\mathcal{O}(\sqrt{n}\log n)$ Sybils per attack edge~\cite{sybilguard}, it suffers from a high false-negative rate (FNR)~\cite{sybil-defenses-survey}. The improved algorithm \textsc{SybilLimit} reduces the number of accepted Sybils per attack edge to $\mathcal{O}(\log n)$ while allowing more attack edges~\cite{sybillimit}.

\textsc{SybilInfer}~\cite{sybilinfer} is a centralized random walk procedure that uses a probabilistic model via Bayesian inference. This allows the algorithm to assign a degree of certainty besides just classifying the nodes.

\textsc{SybilRank}~\cite{sybilrank} uses early-stopping random walks to propagate trust scores rooting from a set of known honest nodes based on the assumption that the honest region of the network is fast-mixing. The trust scores are degree-normalized and ranked, allowing classification with a threshold value. All security guarantees outlined in the paper are based on the assumption that the attack edges are randomly established between honest and Sybil nodes.

Existing random walk methods have the disadvantage that they can only leverage known honest nodes or known Sybil nodes, but not both simultaneously~\cite{sybiledge}. They also tend to lack robustness to label noise in the set of known nodes~\cite{sybilscar}.

% LBP based

\textsc{SybilBelief}~\cite{sybilbelief} utilizes a set of known honest nodes and, optionally, a set of known Sybil nodes to perform classification. Accounting for prior probabilities, the algorithm uses LBP to infer the posterior probabilities of nodes being Sybil.

\textsc{SybilSCAR}~\cite{sybilscar} aims to combine the approaches of random walks and LBP by introducing a novel local update rule which is applied iteratively for a given number of iterations or until convergence. The algorithm has two variants: \textsc{SybilSCAR-C} assumes constant homophily between any two nodes, and takes a parameter specifying this. Setting this parameter, however, usually requires some analysis of the full graph. The other variant is \textsc{SybilSCAR-D} which computes homophily for each edge individually.

\textsc{SybilHP}~\cite{sybilhp} was developed for directed social networks and uses LBP combined with a homophily estimator to classify the nodes. It is designed to overcome the limitation many algorithms have of implicitly assuming constant homophily, which doesn't account for the directional nature of some social trust relationships.

Approaches based on LBP can incorporate knowledge about both known honest and known Sybil nodes simultaneously and tend to be more robust to label noise~\cite{sybiledge}.

\paragraph{Feature-augmented methods}
Increasingly, there has been research that uses features in addition to the graph structure to perform a more accurate classification.

\textsc{TrustGCN} leverages graph convolutional networks (GCNs) to classify nodes using a two-stage process: trust propagation (via random walks) and trust-guided graph convolution~\cite{trustgcn,GCNConv}.

\textsc{BotRGCN} is a Twitter bot detection algorithm that leverages different possible kinds of edges in a social network by applying a relational graph convolutional network (RGCN)~\cite{botrgcn,RGCNConv}.

\textsc{SATAR}~\cite{satar} is a self-supervised representation learning framework for Twitter bot detection. It aims to adapt better to different types of social media bots and is proven to generalize in real-world scenarios.

\paragraph{Early Sybil detection}

Some methods specialize in early Sybil detection, using additional information about friend request targets and responses, along with the network topology. The motivation for these methods is the prevention of Sybils in the network, not just their detection after they have established themselves, aiming to avoid the negative effects they have on the network. \textsc{SybilEdge} aggregates over these features, giving more weight to the request targets that are preferred by other Sybil nodes (in contrast to honest nodes) while considering how these friend request targets respond~\cite{sybiledge}. \textsc{PreAttacK} uses initial friend request behaviors to perform classification by approximating the posterior probability that a new node is Sybil or not under their proposed multi-class Perferential Attachment model for unanswered friend requests~\cite{preattack}. \textsc{PreAttacK} can successfully (AUC $\approx 0.9$) detect Sybil nodes before any edges have been actualized (that is, friend requests have been accepted).

\section{Methodology}

\subsection{Problem Definition}\label{sec:problem-definition}

Given a social network $G=(V,E)$, which consists of honest (negative) nodes $H$ and Sybil\footnote{A Sybil entity is not part of the intended set of entities for a particular network, and is often malicious or disruptive to the collective or individual entities.} (positive) nodes $S$, and small subsets of known nodes $(H_\text{train},S_\text{train})$, we want to perform the Sybil classification. The size of the training (known) nodes is assumed, unless otherwise mentioned, to be 5\% of the respective original node set size.

The edges of these social networks are assumed to be some kind of trust relationship and can be directed or undirected. Since, due to the nature of the available underlying data or for other reasons, much of the previous work focuses on the evaluation of undirected graphs.\footnote{Many online trust relationships are naturally undirected.} We will follow suit in this. Note that a directed network can be transformed into an undirected one by either keeping all edges or only the bidirectional ones while losing the information that the directedness might have implied. Although the edges represent trust relationships, the edges between the honest and Sybil nodes are compromised. These edges are called \emph{attack edges}.

\begin{table}[t]
    \centering
    \begin{tabular}{l|l}
        Description & Notation \\
        \hline 
        Nodes & $V$, $|V|=n$ \\
        Edges & $E$, $|E|=m$ \\
        Honest region & $H$, $|H|=n_H$ \\
        Sybil region & $S$, $|S|=n_S$ \\
        Training (known) nodes & $(H_\text{train},S_\text{train})$ \\
        Test nodes & $(H_\text{test},S_\text{test})$ \\
        Target sets (honest, sybil) & $T_H,T_S$ \\
        Attack edges & $E_T$, $|E_T|=m_T$ \\
        Targeted attack edge probability & $p_T$ \\
        Attack target PDF & $p$ \\
    \end{tabular}
    \caption{Notations.}
    \label{tab:notation}
\end{table}

\paragraph{Why no additional features?}
The problem definition above is designed in a way that takes as input only the graph structure and a small set of known labels. Several recent studies performed Sybil detection using additional features. This could include analyzing a user's posts or other activity in an online social network platform. We purposefully do not include such features for multiple reasons:
\begin{itemize}
    \item With the rise of generative artificial intelligence (GenAI), humans can create human-like entities (e.g., bots, something that would be classified as Sybil) in social networks that become increasingly harder to distinguish from humans--even by humans.
    \item Including additional features always raises concerns about data privacy. By including fewer features (only the structure and some labels), we can circumvent most problems of this kind.\footnote{Of course in some cases, especially when data is sparse, even though the structure alone can allow someone to deduce the identity of some nodes, this must still be addressed.}
    % \item Solving this problem would yield a method that is entirely independent of methods using only node features, and these methods can thus complement each other.
    \item A structure-based approach offers enhanced generalizability across diverse social platforms and cultural contexts, as network patterns often remain consistent even when user behaviors and content vary significantly.
\end{itemize}

\subsection{Social Network Synthetization}\label{sec:social-network-synthetization}

Due to limited access to labeled social network datasets, previous research resorted to evaluating (and developing) their algorithm on synthesized social networks.\footnote{We, just like the authors of some previous work, acknowledge this limitation and the effects thereof.} 
We will adopt the methods and parameters from related research~\cite{sybilguard,sybilinfer,sybillimit,sybilrank,sybildefender,sybilframe,sybilwalk,sybiluncover,sybilscar,sybilfence,sybilexposer}.

The general approach assumes that social networks consist of honest and Sybil regions. Following previous research, and to allow proper comparison, we will assume that there is one of each region, but this approach can be generalized to allow multiple regions.

We construct a synthetic social network as follows: We take as honest and Sybil regions real-world social network graphs or synthesized graphs (\Cref{sec:synthetic-regions}). %~\cite{sybildefender,sybilframe,sybilwalk,sybilscar}.
These regions are assumed to be tightly connected within and are connected by a certain number of attack edges. We measure the number of attack edges in the unit of attack edges per Sybil, that is, the average number of edges that cross over the regions per Sybil. Of course, the more attack edges there are, the harder the problem becomes, since then the regions no longer present themselves as tightly connected as they initially were and start to ``blend'' with one another (e.g., the Sybils can ``convince'' many honest nodes to engage in trust relationships with them, making themselves appear more honest according to the homophily assumption). Increasing the number of attack edges is a common way to make the problem harder and can show distinguishable performance differences between algorithms.

This methodology for social network synthetization is prevalent in most previous research in structure-based Sybil detection.%~\cite{sybilguard,sybilinfer,sybillimit,sybilrank,sybildefender,sybilframe,sybilwalk,sybiluncover,sybilscar}.

\subsubsection{Synthetic Regions}\label{sec:synthetic-regions}

The following synthetic models\footnote{The implementations of the synthetic graph generators used are from \texttt{networkx}~\cite{networkx}.} were chosen for our evaluation. These models are used in related works, and the parameters were inspired by previous research in conjunction with the analysis of real social networks. %~\cite{sybilrank,}.
%These models are used in related works, and the default parameters were inspired by past research in conjunction with analysis of real social networks~\cite{sybilinfer,sybilrank,sybildefender,sybilfence,sybilframe,sybilexposer,sybiluncover}.

%\paragraph{Erdős–Rényi (ER) Model}
%The Erdős–Rényi (ER) generator creates a $G_{n,p}$ (Erdős–Rényi or binomial) graph with n nodes by choosing each of the possible edges with probability p~\cite{erdos-renyi}. We set these parameters to $(n,p)=(\cdot,6/n)$ by default.

\paragraph{Barabási–Albert (BA) Model}
Barabási–Albert (BA) model generator creates a graph of n nodes by attaching new nodes (with $m\in\{1,\ldots,n\}$ edges each) that are preferentially attached to existing nodes with high degree~\cite{barabasi-albert}. Our standard parameter configuration is $(n,m)=(\cdot,6)$.

\paragraph{Power law (PL) Model}
The Power law (PL) graph generator is an algorithm for generating graphs with power law distributed degrees, and approximate average clustering~\cite{power-law}. The parameters of this random graph generator are $n$ (the number of nodes), $m\in\{1,\ldots,n\}$ (the number of random edges to add for each node), and $p\in \left[0,1\right]$ (probability of adding a triangle after adding a random edge). Unless otherwise mentioned, these parameters are set to $(n,m,p)=(\cdot,6,0.8)$. This is equivalent to the BA model, but with the added chance (controlled by $p$) that a newly added random edge is closed to form a triangle~\cite{barabasi-albert}. A graph generated with the PL model may be disconnected (but this did not happen in our experiments).

\subsubsection{Attack Edge Placement}

The placement of these synthesized attack edges can be performed with two general strategies: \emph{random} or \emph{targeted}. The attack edges $E_T$ are placed between honest and Sybil (sub)sets, $T_H\subseteq H$ and $T_S\subseteq S$.

\paragraph{Random attack edges}
These edges are placed uniformly at random between nodes of the honest and Sybil target sets, which are set to $T_H=H$ and $T_S=S$.

\paragraph{Targeted attack edges}
In our methodology for targeted attack edges, an attacking entity can establish target sets. Unless otherwise stated, these sets are assumed to be $T_H=H_\text{train}$ and $T_S=S$, which means that all Sybil nodes attempt to target the set of known honest nodes. Generally, it is assumed that a targeted attack edge ``originates'' from a Sybil node and targets an honest node, following the notion that the Sybil nodes aim to disturb the system. A targeted attack further consists of two parameters, a targeted attack edge probability $p_T\in[0,1]$ and a discrete probability distribution function $p\in[0,1]^{K+1}$ ($\sum_{k} p_k=1$). The parameter $p_T$ describes the probability that a targeted attack edge is placed (otherwise a random attack edge is placed between $S$ and $H$). 
The implicitly defined $K$ is the maximum distance from a node in $T_H$ that a Sybil node $u$ hits with a targeted attack. 
% For each attack edge $e=(u,v)\in E_T$ originating from a randomly chosen $u\in T_S$, given that such a targeted attack occurs, the discrete PDF $p$ and the implicitly defined maximal hit distance $K$ specifies the probability distribution of how close to the target set a given targeted attack edge will hit. 
Formally, this means that for the attack edge $e=(u,v)$, given a targeted attack edge is placed, it holds that
\begin{equation}
    p_k=\mathbb P\left[v\in D_k(T_H)\right]\text{ for }k\in\{0,\dots,K\}.
\end{equation}
Here, $D_k(\cdot)$ denotes the set of nodes distance $k$ from any node in $T_H$ ($k=0$ denotes a ``direct hit''). 
%it holds that $v=v'$, since $v'$ is the only node in the 0-hop neighborhood of $v'$ itself (this is a ``direct hit'').

A random attack edge strategy is equal to a targeted strategy where $p_T=0$ (making $p$ irrelevant) since all edges will end up being random. Of course, a randomly placed attack edge can coincidentally be the same as a targeted edge.

In summary, for a general attack with $m_T$ attack edges and parameters $p_T$ and $p$ as above, the expected number of random attack edges is $\mathbb E[|E_{T_\text{rand}}|]=(1-p_T) m_T$, and the expected number of attack edges hitting a $k$-hop neighbor of a node in the specified Sybil target set is $\mathbb E[|E_{T_\text{targ}}^{(k)}|]=p_T \cdot p_k\cdot m_T$. 
%The following equations show that this setup of targeted attack ends up with an expected (and enforced by our code implementation) number of attack edges $m_T$.

%\begin{align*}
%    \mathbb{E}[m_T]&=\mathbb{E}[m_{T_\text{rand}}+m_{T_\text{targ}}] \\
%    &=\mathbb{E}[m_{T_\text{rand}}]+\mathbb{E}[m_{T_\text{targ}}] \\
%    &=\mathbb{E}[m_{T_\text{rand}}]+\sum_{k=0}^K\mathbb{E}[m_{T_\text{targ}}^{(k)}] \\
%    &=(1-p_T) m_T+\sum_{k=0}^K p_T \cdot p_k\cdot m_T \\
%    &=(1-p_T) m_T+p_T\cdot m_T\underbrace{\sum_{k=0}^K p_k}_{=1}=m_T
%\end{align*}

\subsection{Graph Neural Network Model}

Our evaluation of the feasibility of using Graph Neural Networks (GNNs)\footnote{The implementation for different graph neural network layers were taken from \texttt{torch-geometric}~\cite{torch-geometric}.} to detect Sybils in online social networks involved different GNN architectures.

% Old version
\emph{Graph Convolutional Networks (GCNs)} aggregate node neighbors, and feed the aggregated features through a traditional neural network, which are then taken as the features in the next layer~\cite{GCNConv}. The parameters of this neural network make up the parameters of the GNN, where each edge is treated equally. The values are propagated through the layers of the network. Using GCNs to perform Sybil detection can keep up with many of the compared baselines, but does not consistently outperform them.

% New version
%\emph{Graph Convolutional Networks (GCNs)}~\cite{GCNConv} were evaluated for Sybil detection. This approach can keep up with many of the compared baselines, but does not consistently outperform them.

%\emph{Relational Graph Convolutional Networks (RGCNs)} are another type of GNN, which work in the same way as GCNs, with the difference that in RGCNs, different edge types can be specified, and will have different parameters in the neural network~\cite{RGCNConv}. Using RGCNs in Sybil detection can be superior when the underlying social network has many targeted attack edges. There are nine possible configurations any given attack edge $(u,v)$, namely
%\begin{equation}
%    (u,v)\in\{\underbrace{\text{known honest}}_{H_\text{train}},\underbrace{\text{known Sybil}}_{S_\text{train}},\underbrace{\text{unknown}}_{H_\text{test}\cup S_\text{test}}\}^2.
%\end{equation}
%Up to symmetry, this results in six different edge types that the RGCN is made up of, allowing the RGCN to learn different parameters depending on what kind of nodes the edges connect. This thus gices the model more predictive power than the GCN. The downside of the RGCN setup is that it performs notably worse when there are mostly random attack edges, making the different targeted edge types too infrequent.

\emph{Relational Graph Convolutional Networks (RGCNs)} are potentially interesting for Sybil detection, as they allow the specification of different types of edges, and learn parameters according to this distinction~\cite{RGCNConv}. This could be beneficial for targeted attacks, where the different types of edges could play a more significant role. When choosing the types of edges to be the different possible combinations between known honest and Sybil nodes, and unknown nodes,
% *namely
%\begin{equation}
%    (u,v)\in\{\underbrace{\text{known honest}}_{H_\text{train}},\underbrace{\text{known Sybil}}_{S_\text{train}},\underbrace{\text{unknown}}_{H_\text{test}\cup S_\text{test}}\}^2,
%\end{equation}
%resulting in (up to symmetry) six different configurations, 
the RGCN performs well in a targeted attack setting. However, when the attack edges are predominantly random, the RGCN's performance is much worse.

\emph{Graph Attention Networks (GATs)} introduce an attention mechanism to assign different weights to different nodes in the neighborhood~\cite{GATConv}. This can allow the model to focus on certain nodes during aggregations. The attention mechanism operates over neighborhoods, and unlike in ``vanilla'' GNNs which have globally learned weights, GATs assign different, learnable weights to neighbors dynamically, which might be interesting to Sybil detection where some nodes (Sybils) disrupt the network by infiltrating it with attack edges. This approach works well and is the basis of our algorithm, presented in \Cref{sec:sybilgat-algorithm}.

\subsubsection{Why GNNs for Sybil Detection?}

Among previous work on structure-based Sybil detection, there is a common trend of using approaches based on random walks (RW) and loopy belief propagation (LBP). These methods constitute the current state-of-the-art for this problem. Due to the nature of GNNs, they should -- at least in theory -- be just as powerful as LBP or RWs. The motivation for using GNNs for Sybil detection is not only this potential theoretical superiority, but it also presents the opportunity to no longer have to rigorously analyze different graphs to arrive at an algorithm design that might apply to only certain scenarios but to design and test different GNN architectures that can self-adapt -- taking out the guesswork while generalizing well.

\subsubsection{Pre-training on Smaller Graphs}
The main mechanism used in this work to perform Sybil detection with GNNs is to run the algorithm on a known small network graph (e.g., a sampled subgraph of the social network graph of interest), and then apply the model to a larger network graph, where only a small number of nodes are known (e.g., the remaining network graph after said subsampling). The performance is then evaluated solely on the evaluation network graph. We will consider the evaluation network to be disjunct from the initial pre-training network, to make for a more realistic setting and more fair evaluation.

\subsubsection{Transductive Learning}
Another way Sybil detection can be performed on social network graphs using GNNs is through \emph{transductive learning}. In this setting, the GNN algorithm does not perform separate pre-training, but runs on the social network graph of interest directly, with a small set of known (train) nodes, and concludes by predicting labels for all nodes of the graph. The prediction is then evaluated on the set of unknown (test) nodes.\footnote{Some past research papers have evaluated their algorithm on all nodes, even the known ones. We do not do this.} We will focus on the aforementioned pretraining approach, and only use transductive learning in \Cref{sec:experiment4}.

\subsection{\textsc{SybilGAT}: Detecting Sybils with GATs}\label{sec:sybilgat-algorithm}

In this work, we present \textsc{SybilGAT}, a GNN algorithm for Sybil detection based on the \texttt{GATConv} layer~\cite{GATConv}.

\subsubsection{Model Parameters} The model itself consists of the following hyperparameters: input width, hidden width, output width (number of classes), number of attention heads, and number of layers (the depth). The input width can be either 1 (representing ``Sybil-ness'') or 2 (a channel for honest and one for Sybil), where we found the former to be conceptually simpler and at least as effective. The hidden width and number of heads are hyperparameters that can be heavily experimented with. We ended up with a robust evaluation by taking both the hidden width and the number of heads as 4. Like the input, the output width can be 1 or 2. The output is then used for prediction depending on some threshold. The number of layers dictates how far into the network layers are aggregated, and an optimal value depends on the structure of the network.

\subsubsection{Model Architecture}
Suppose the model has parameters $I$ for input width, $H$ for hidden width, $O$ for output width, $N$ for number of heads, and $L$ layers. The first layer is a \texttt{GATConv} layer with $I$ input units and $H$ output units, with $N$ heads. The intermediate layers have $H\cdot N$ input units, $H$ output units, and $N$ heads. The last layer has $H\cdot N$ input units, $O$ output units, and 1 head. Before each layer, there is a dropout layer with a probability of $0.5$. After each layer, there is a tanh activation layer. After the last layer, there is a sigmoid ($O=1$) or softmax ($O=2$) activation function.

\subsubsection{Training Procedure with Early Stopping}
Initially, the training set of known nodes is split into an actual training set and a validation set used for early stopping with a specifiable patience parameter. By default, the train/validation split is 0.8/0.2 for the training phase and 0.9/0.1 for the inference (prediction) phase. A set of known labels are used as inputs (depending on the input width) and fed through the network. The predictions made by the model are compared with the ground-truth label output of the known nodes. To this, a loss function is applied, and an optimizer performs the backward step. In our experiments, we used the binary cross entropy loss for ($O=1$), the cross entropy loss ($O=2$), and the Adam optimizer. If there has been no improvement in validation loss for the last epochs (specified by the patience parameter), the training process is stopped and the best model (according to validation loss) is retrieved for prediction. The predictions are then evaluated in terms of some metric on the test set (the remaining nodes).

\subsubsection{Prediction Threshold Estimation}
During inference and before prediction, a threshold value is computed. This is done using the 10\% of known nodes in the validation set, as mentioned above. The optimal threshold is computed for the validation set and this estimate is used for prediction.

\subsection{Sampling Subgraphs of Social Networks}

The sampling method used in the evaluation is the \emph{Forest Fire} sampling method~\cite{FF-sampling-details,FF-sampling}. It implements\footnote{The \texttt{little-ball-of-fur}~\cite{little-bar-of-fur} Python library was used for the graph subsampling.} a stochastic snowball sampling method with a specifiable burning probability that is proportional to the expansion~\cite{little-bar-of-fur}. 

\section{Experimental Results}

\subsection{Setup}

\subsubsection{Datasets}

The Twitter\footnote{\url{https://twitter.com}. Twitter is now named X, \url{https://x.com}. Since all research and datasets considered are from before this name change, we will refer to the platform as Twitter.} dataset is a real-world social network graph consisting of 269'640 nodes and 6'818'501 edges. The nodes represent users, and the directed edges represent the ``following'' relationship. Before evaluation, this graph is transformed into an undirected graph. This dataset was sampled and processed~\cite{sybilhp} from a previously much larger crawled graph~\cite{what-is-twitter}. Initially, the Twitter API was used to crawl the graph and then, retroactively and repeatedly over the past few years, determine which accounts were honest or Sybil accounts by querying their account status.

The Facebook graph~\cite{snap-facebook-dataset,snap-datasets} from SNAP\footnote{\url{https://snap.stanford.edu/index.html}.} is an undirected, unlabeled social network graph with 4'039 nodes and 88'234 edges. The dataset is a friendship network from Facebook, where the nodes are users and the edges are friendships between the users. Following previous research~\cite{sybildefender,sybilframe,sybilwalk,sybilscar}, we will be using this graph as regions of a synthesized social network. Since this graph is very highly connected, the synthesized network is created with a high number of attack edges (in our evaluation, 20 attack edges per Sybil).

In the following experiments, we will use both these two real-world data sets (either directly or as real regions), as well as fully synthesized social networks (cf. \Cref{sec:social-network-synthetization}).

\subsubsection{Baseline Algorithms}

From the list of previous research that study Sybil detection using only the network structure (Cf. problem definition in \Cref{sec:problem-definition}) we narrowed our baselines to three algorithms which have consistently been used as baselines: \textsc{SybilRank}, \textsc{SybilBelief} and \textsc{SybilSCAR}~\cite{sybilrank,sybilbelief,sybilscar}. The latter is used in its $D$ variant due to its flexibility and the lack of need for analysis of the full graph, as described in \Cref{sec:related-work}. \textsc{SybilSCAR}, the most recent of them, consistently outperforms the other baseline algorithms and is more robust in different evaluation scenarios. For this reason, it is our main baseline and is used for the first three experiments. A comparison between all baseline algorithms with varying numbers of attack edges can be found in Experiment 4 in \Cref{sec:experiment4}. Full tables for all experiments with results for all algorithms can be found in the appendix. For the implementation of \textsc{SybilSCAR}, we used the matrix-form algorithm described by the authors and optimized its runtime by using sparse matrix operations (allowing it to run in matrix form when evaluating large graphs such as the Twitter network). We tested against the public C++ code by the authors, and our implementation performed equally (up to numerical differences, and sometimes better) to the comparison. Due to this, we used our implementation.

\subsubsection{Experiments}

In the following four sections, we will present the results of our experiments:

\begin{enumerate}
    \item Pre-training on Sampled Subgraph (\Cref{sec:experiment1})
    \item Pre-training on Small Synthetic Network (\Cref{sec:experiment2})
    \item Attacking after Pre-training (\Cref{sec:experiment3})
    \item General Robustness Baseline Comparison (\Cref{sec:experiment4})
\end{enumerate}

Each experiment is performed five times,\footnote{The seeds for the five experiments are $[42,43,44,45,46]$.} and the mean is calculated. The performance metric we will focus on is the AUC (Area under the ROC curve) score. For most experiments, we will evaluate three instances of \textsc{SybilGAT}: a shallow, intermediate, and deep model with 2, 4, and 8 layers, respectively.

\subsection{Experiment 1: Pre-training on Sampled Subgraph}\label{sec:experiment1}

Using the sampling method described above, we will produce a subgraph of a social network, which will be used by \textsc{SybilGAT} for pretraining. The evaluation (prediction) will then be performed exclusively on the remaining graph.

For all experiments, the size of the subgraph is 10\% of the initial graph, except for the Twitter graph, where it is 5\%. The training set for the Twitter graph is 11.2\% (honest) and 10.9\% (Sybil) of the respective total sizes~\cite{sybilhp}. 

\subsubsection{Real Twitter Dataset}

Using the Twitter dataset introduced above, we evaluated the performance of \textsc{SybilGAT} by training on a subset of the graph using the forest fire sampling method.
The results seen in \Cref{tab:experiment1-results-twitter-facebook} show that the best \textsc{SybilGAT} instance performs up to 5\%-points better than \textsc{SybilSCAR}.

\subsubsection{Synthesized Social Network with Real Honest and Sybil Regions}

Here, we used the Facebook graph as the honest and Sybil regions of the graph and added 20 (random) attack edges per Sybil to create the network. The two strategies used are random placement, and targeted placement with attack probability $p_T=0.1$ and discrete target hit distance PDF $p=[\frac{1}{4},\frac{1}{4},\frac{1}{2}]$.

The results in \Cref{tab:experiment1-results-twitter-facebook} show superior results for the shallow version $\textsc{SybilGAT-L2}$. The deep model of \textsc{SybilGAT} performs very poorly, most likely due to the high average degree of the Facebook graph, omitting the need for propagating values very far through the network, essentially rendering the deep model too complex.

\begin{table}[t]
    \centering
    \begin{tabular}{l|c|cc}
        \textbf{Algorithm} & \multicolumn{3}{c}{\textbf{AUC}} \\
        \hline 
        Dataset & Twitter & \multicolumn{2}{c}{Facebook (synth.)} \\
        \hline 
        Attack edges & N/A & random & targeted \\
        \hline 
        \textsc{SybilSCAR} & 0.8022 & 0.6265 & 0.5029  \\
        \textsc{SybilGAT-L2} & 0.8254 & \textbf{0.7479} & \textbf{0.7739} \\
        \textsc{SybilGAT-L4} & \textbf{0.8489} & 0.5980 & 0.5942 \\
        \textsc{SybilGAT-L8} & 0.7973 & 0.4474 & 0.4282 \\
    \end{tabular}
    \caption{Results for experiment 1. The attack edges are the type of attack edges that are added to the synthesized network (not applicable to the real-world Twitter dataset).}
    \label{tab:experiment1-results-twitter-facebook}
\end{table}

\subsubsection{Fully Synthesized Network}

For the fully synthesized network we evaluate two sizes of networks: 10'000 nodes and 50'000 nodes. Both networks are created with the power law model with parameters $m=5$ and $p=0.8$, and 8 (random) attack edges are added per Sybil.

The results in \Cref{tab:experiment1-results-synthetic} indicate that \textsc{SybilGAT-L4} achieves the highest score in both inspected networks. Also, given that both networks produce almost identical scores for each algorithm, the network size is not a relevant factor given a certain synthetization scheme.

\begin{table}[b]
    \centering
    \begin{tabular}{l|c|c}
        \textbf{Algorithm} & \multicolumn{2}{c}{\textbf{AUC}} \\
        \hline 
        \# nodes & 10'000 & 50'000 \\
        \hline 
        \textsc{SybilSCAR} & 0.5574 & 0.5565 \\
        \textsc{SybilGAT-L2} & \textbf{0.6219} & \textbf{0.6237} \\
        \textsc{SybilGAT-L4} & 0.5941 & 0.5960 \\
        \textsc{SybilGAT-L8} & 0.5511 & 0.5531 \\
    \end{tabular}
    \caption{Results for experiment 1 on fully synthesized social networks, using the power law (PL) model.}
    \label{tab:experiment1-results-synthetic}
\end{table}

\subsection{Experiment 2: Pre-training on a Smaller Synthetic Social Network}\label{sec:experiment2}

In this experiment, instead of pre-training on an actual subgraph of a large network, \textsc{SybilGAT} is pre-trained on a smaller version of the synthesized network before being applied to a larger network with the same underlying model (exception is the last case, where we apply it to a network synthesized using the Facebook graph--a scenario we consider useful and close to the real world).

Three cases were evaluated: the synthetic models Barabási–Albert (BA) and power law (PL), and pre-training on a small synthesized power law network before applying to a synthesized network with the Facebook graph as real regions (PL-FB). In each experiment, the small network consists of 2000 nodes, and the large network consists of 20'000 nodes (except the Facebook network, where the size is given by the underlying real graph--namely 8'078 nodes). The network is filled with 8 attack edges per Sybil (20 for the Facebook evaluation), either randomly or targeted ($p_T=0.1$, $p=[\frac{1}{4},\frac{1}{4},\frac{1}{2}]$).

The results in \Cref{tab:experiment2-results} show that for the Barabási–Albert model, \textsc{SybilGAT-L8} outperforms the other algorithms notably. In the power law model, all \textsc{SybilGAT} instances are very similar in performance while significantly outperforming \textsc{SybilSCAR}. In the last experiment, which evaluates the synthesized Facebook network, \textsc{SybilGAT-L2} achieves the highest, while the deep model performs very poorly for the same reason as mentioned above.

\begin{table}[t]
    \centering
    \begin{tabular}{cl|cc}
        & \multicolumn{1}{c|}{\textbf{Algorithm}} & \multicolumn{2}{c}{\textbf{AUC}}\\
        \cline{2-4}
        & \multicolumn{1}{c|}{Attack edges} &  \multicolumn{1}{c}{random} & \multicolumn{1}{c}{targeted}\\
        \cline{2-4}
        \parbox[t]{2mm}{\multirow{4}{*}{\rotatebox[origin=c]{90}{BA-BA}}} & \textsc{SybilSCAR} & 0.6740 & 0.4686 \\
        & \textsc{SybilGAT-L2} & 0.7506 & \textbf{0.6021} \\
        & \textsc{SybilGAT-L4} & 0.8068 & 0.5586 \\
        & \textsc{SybilGAT-L8} & \textbf{0.8589} & 0.4171 \\
        \cline{2-4}
        \parbox[t]{2mm}{\multirow{4}{*}{\rotatebox[origin=c]{90}{PL-PL}}} & \textsc{SybilSCAR} & 0.6647 & 0.4775 \\
        & \textsc{SybilGAT-L2} & 0.7402 & \textbf{0.6262} \\
        & \textsc{SybilGAT-L4} & \textbf{0.7565} & 0.5604 \\
        & \textsc{SybilGAT-L8} & 0.7428 & 0.4681 \\
        \cline{2-4}
        \parbox[t]{2mm}{\multirow{4}{*}{\rotatebox[origin=c]{90}{PL-FB}}} & \textsc{SybilSCAR} & 0.6458 & 0.5199 \\
        & \textsc{SybilGAT-L2} & \textbf{0.7682} & \textbf{0.7795} \\
        & \textsc{SybilGAT-L4} & 0.6105 & 0.5742 \\
        & \textsc{SybilGAT-L8} & 0.4746 & 0.4441 \\
    \end{tabular}
    \caption{Results for experiment 2. BA-BA: Barabási–Albert, PL-PL: Power law, PL-FB: Power law and Facebook.}
    \label{tab:experiment2-results}
\end{table}

\subsection{Experiment 3: Attacking the Social Network after Pre-training}\label{sec:experiment3}

In this experiment, \textsc{SybilGAT} is pre-trained on a social network that was attacked with 8 random attack edges per Sybil (20 for the Facebook evaluation). It is then evaluated on a social network consisting of identical honest and Sybil regions, but attacked with 8 attack edges per Sybil (20 for the Facebook evaluation) following the targeted attack parameters $p_T=0.2$ (20\% of attack edges will be targeted, the rest will be random) and the discrete target hit distance PDF $p=[\frac{1}{2},\frac{1}{2}]$ (half of all targeted edges will hit a known node directly, the other half will hit a neighbor). The sizes of the social networks (except for the one involving the Facebook graph) are 2000 nodes.

\subsubsection{Synthesized Network with Real Honest and Sybil Regions}

In this experiment, the Facebook graph was used as the honest and Sybil region. \Cref{tab:experiment3-results-facebook} shows, similarly to \Cref{tab:experiment1-results-twitter-facebook} (which also inspected the Facebook network), that \textsc{SybilGAT-L2} outperforms the other algorithms. As described previously, \textsc{SybilGAT-L8} performs very poorly.

\begin{table}[t]
    \centering
    \begin{tabular}{l|c}
        \textbf{Algorithm} & \textbf{AUC} \\
        \hline 
        \textsc{SybilSCAR} & 0.6422 \\
        \textsc{SybilGAT-L2} & \textbf{0.7669} \\
        \textsc{SybilGAT-L4} & 0.6019 \\
        \textsc{SybilGAT-L8} & 0.4690 \\
    \end{tabular}
    \caption{Results for experiment 3 on a synthesized social network with the Facebook graph.}
    \label{tab:experiment3-results-facebook}
\end{table}

\subsubsection{Fully Synthesized Network}

In this part, the two random graph models Barabási–Albert (BA) and power law (PL) were used to generate synthetic social networks.
The scores in \Cref{tab:experiment3-results-synthetic} show that in the BA model, \textsc{SybilGAT-L8} significantly outperforms \textsc{SybilSCAR}. Using the PL model \textsc{SybilGAT-L4} has the best score.

\begin{table}[t]
    \centering
    \begin{tabular}{cl|c}
        & \multicolumn{1}{c|}{\textbf{Algorithm}} & \multicolumn{1}{c}{\textbf{AUC}}\\
        \cline{2-3}
        \parbox[t]{2mm}{\multirow{4}{*}{\rotatebox[origin=c]{90}{BA}}} & \textsc{SybilSCAR} & 0.6714 \\
        & \textsc{SybilGAT-L2} & 0.7326 \\
        & \textsc{SybilGAT-L4} & 0.7788 \\
        & \textsc{SybilGAT-L8} & \textbf{0.8273} \\
        \cline{2-3}
        \parbox[t]{2mm}{\multirow{4}{*}{\rotatebox[origin=c]{90}{PL}}} & \textsc{SybilSCAR} & 0.6367 \\
        & \textsc{SybilGAT-L2} & 0.7087 \\
        & \textsc{SybilGAT-L4} & \textbf{0.7147} \\
        & \textsc{SybilGAT-L8} & 0.6836 \\
    \end{tabular}
    \caption{Results for experiment 3 on fully synthesized social networks. BA: Barabási–Albert, PL: Power law.}
    \label{tab:experiment3-results-synthetic}
\end{table}

\subsection{Experiment 4: General Robustness Evaluation}\label{sec:experiment4}

We evaluate the general robustness of the algorithms,\footnote{\textsc{SybilGAT-L8} was not evaluated in this experiment since its performance was unstable in previous experiments.} including \textsc{SybilRank} and \textsc{SybilBelief}.
The experiment is set up as follows: Synthetic social networks are created using the Barabási–Albert (BA) and power law (PL) models to have a size of 2000 nodes. The number of attack edges per Sybil ranges from 1 to 12, which represents an increasing difficulty of the problem given random attack edges.

\begin{figure}[t]
    \centering
    \includegraphics[width=\linewidth]{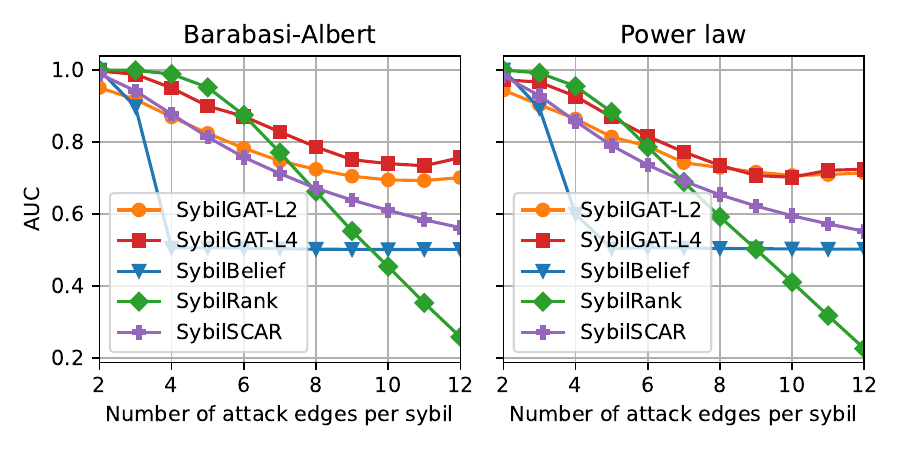}
    \caption{AUC score plots for experiment 4 with the Barabási–Albert (BA) and Power law (PL) models, varying number of attack edges per Sybil.}
    \label{fig:experiment4-results}
\end{figure}

\Cref{fig:experiment4-results} shows the AUC score of the inspected algorithms while increasing the attack edges per Sybil. The plots show clearly that, with increasing attack edges per Sybil, the \textsc{SybilGAT} algorithms outperform the baselines, especially when the problem gets very hard.

% \begin{figure}[t]
%     \centering
%     \includegraphics[width=\linewidth]{exp4_PL.pdf}
%     \caption{AUC score plot for experiment 4 with the power law (PL), varying number of attack edges per Sybil.}
%     \label{fig:experiment4-results-PL}
% \end{figure}

\section{Conclusion}

This paper introduced \textsc{SybilGAT}, a novel approach for Sybil detection using Graph Attention Networks. Our experiments demonstrated that \textsc{SybilGAT} consistently outperforms the state-of-the-art algorithms in various types of networks and attack scenarios. Key findings include superior performance on both real-world and synthetic datasets, effective pre-training on smaller networks for application to larger ones, and maintained performance under targeted attacks. We show that the method can be applied to a real-world graph from Twitter with 269k nodes and 6.8M edges.  

%The robust performance of \textsc{SybilGAT}, especially as the complexity of the attacks increases, represents a significant advancement in the security of social networks. However, the performance of the depth of the model showed some variation between the network structures, indicating the need to evaluate a model before deploying deep instances.
The robust performance of \textsc{SybilGAT}, especially as the complexity of the attacks increases, represents a significant advancement in the security of social networks. However, limitations exist: The depth of the optimal model varies with network structures, and its scalability to larger networks and robustness against adversarial attacks remains to be fully explored. These challenges indicate the need for adaptive architectures and further investigation of the performance of \textsc{SybilGAT} on dynamic networks.

% Although limitations naturally exist, \textsc{SybilGAT}'s success opens new avenues for applying graph learning techniques to network security challenges. Future work could explore larger-scale networks and adaptive architectures, focus on gathering more real-world data for testing, or explore why different network structures resulted in different optimal numbers of layers for \textsc{SybilGAT}. In general, \textsc{SybilGAT} offers a promising tool for maintaining the integrity of social networks in the face of evolving Sybil threats that depend solely on the network structure.
\textsc{SybilGAT}'s success opens new avenues for applying graph learning techniques to network security challenges. Future work could address the identified limitations, explore larger-scale networks, focus on gathering more real-world data for testing, and investigate why different network structures result in different optimal numbers of layers. Overall, \textsc{SybilGAT} offers a promising tool for maintaining the integrity of social networks in the face of evolving Sybil threats that depend solely on the network structure.

% Stuart: According to AAAI author kit, appendix must come after main content but before Acknowledgements and References
%\appendix

%\section*{Acknowledgments}

%todo: nothing?

\bibliography{main}

% Stuart: No content allowed after References

\end{document}